\documentclass[conference]{IEEEtran}
\usepackage{geometry}
\IEEEoverridecommandlockouts
\usepackage{cite}
\usepackage{amsmath,amssymb,amsfonts}
\usepackage{algorithmic}
\usepackage{graphicx}
\usepackage{textcomp}
\usepackage{xcolor}
\def\BibTeX{{\rm B\kern-.05em{\sc i\kern-.025em b}\kern-.08em
    T\kern-.1667em\lower.7ex\hbox{E}\kern-.125emX}}
\begin{document}

\title{Deep Reinforcement Learning Control for Disturbance Rejection in a Nonlinear Dynamic System with Parametric Uncertainty
}
\author{\IEEEauthorblockN{Vincent W. Hill}
\IEEEauthorblockA{\textit{The University of Alabama}}
Tuscaloosa, AL, USA
}
\maketitle

\begin{abstract}
This work describes a technique for active rejection of multiple independent and time-correlated stochastic disturbances for a nonlinear flexible inverted pendulum with cart system with uncertain model parameters. The control law is determined through deep reinforcement learning, specifically with a continuous actor-critic variant of deep Q-learning known as Deep Deterministic Policy Gradient, while the disturbance magnitudes evolve via independent stochastic processes. Simulation results are then compared with those from a classical control system.
\end{abstract}


\section{Introduction}
One of the most challenging tasks in the design of a new system is dynamic modeling for simulation. This is especially true for complex vehicles with highly nonlinear dynamics. Specifically, flexible dynamics have become more critical in recent years as a result of the drive to improve vehicle performance by decreasing structural weight. These systems are substantially more difficult to precisely model without undertaking costly and time-consuming ground vibration tests, so numerical studies must be used to capture the effects of flexibility on vehicle dynamics. While these techniques have improved in the last two decades, they are limited in scope and are often only used temporarily until flight data can be gathered.

Added to the challenge of flexible nonlinear dynamic system modeling is prohibitively expensive prototype testing, especially in the case of large aircraft and launch vehicles. Despite these difficulties, high-performance control systems must still be developed in time for the vehicle’s initial deployment. This means control system design must be accomplished with models carrying a high degree of uncertainty, particularly in structural flexibility. 

In addition to potential for uncertainty in vehicle mathematical modeling, aircraft and spacecraft are subject to random excitations arising from atmospheric conditions. These disturbances can be mitigated through passive means, such as fins for launch vehicles or static stability for an aircraft, but these passive disturbance alleviation methods are limited in effectiveness. Active disturbance rejection techniques can handle a broad class of excitation without impacting vehicle performance. 

These factors place a premium on control systems that are robust to both a high degree of dynamic model uncertainty as well as stochastic disturbances while maintaining high performance. Deep Reinforcement Learning (DRL) is a new technology that has emerged in the last several years is well-suited to tackle this difficult problem. Mnih et al. \cite{dqn} introduced the Deep Q Network (DQN) algorithm in 2015 as a method for achieving human-level performance playing Atari games such as \textit{Breakout} by coupling the Q-learning algorithm with deep neural networks for action-value function approximation. DQN operates by recursively updating the Q-function using dynamic programming, and actions are chosen by taking the $argmax$ of $Q$ conditioned on the current state. Over time the DQN algorithm, known as an agent, will optimize this function to determine the best action for every possible state.

DQN is a very effective solution for low-dimensional discrete action spaces such as an Atari controller, but struggles in a continuous action space such as the gimbal angle of a launch vehicle thrust vector controller or the deflection angle of an aircraft elevator. Lillicrap et al. \cite{ddpg} proposed an extension to DQN called Deep Deterministic Policy Gradient (DDPG) that utilizes an actor-critic algorithm with deep function approximators to learn policies in high-dimensional continuous action spaces such as those in a control design problem.

To demonstrate the effectiveness of deep reinforcement learning control for active disturbance rejection in nonlinear systems with parametric uncertainty, the technique is applied to a flexible inverted pendulum with cart (FIPWC) mathematical model with parametric uncertainty and stochastic disturbances. In this work, DRL is used to learn the nonlinear control signal that will keep the pendulum vertical throughout the simulation while under the influence of time-correlated stochastic disturbances. Results from that control system are then compared to results from the use of a conventional proportional-derivative (PD) control law.

\section{Technical Approach}

\subsection{Flexible Inverted Pendulum with Cart System Modeling}

\begin{figure}[htbp]
\centerline{\includegraphics[width=0.4\textwidth]{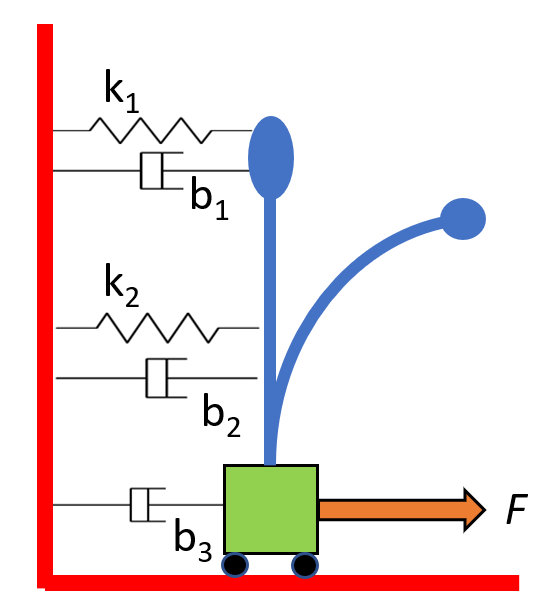}}
\caption{FIPWC system diagram.}
\label{diag1}
\end{figure}

\begin{figure}[htbp]
\centerline{\includegraphics[width=0.4\textwidth]{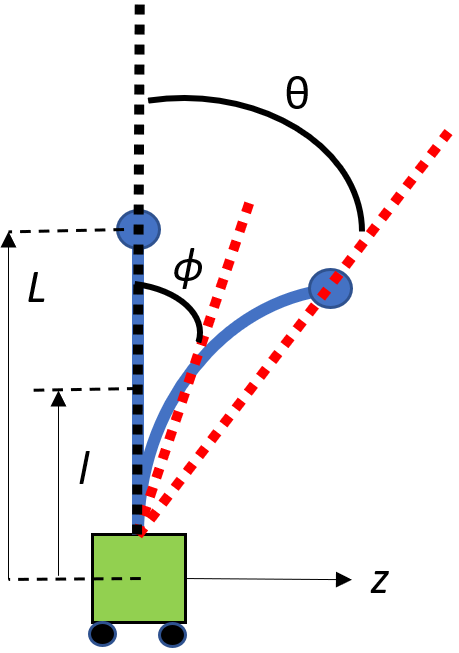}}
\caption{FIPWC system diagram.}
\label{diag2}
\end{figure}

The FIPWC system is modeled as a beam with a tip mass, with flexibility effects modeled as simple linear springs attached at the pendulum tip and center of mass. Figures \ref{diag1} and \ref{diag2} provide a visual overview of the FIPWC system.

The FIPWC mathematical model used in this work was developed by Gorade et al. \cite{flex1} through a multi-body approach with Lagrangian equation of motion derivation. Equation \ref{lagrange} provides the Lagrangian of the system, where $m_t$, $m_b$, and $m_c$ are the masses of the pendulum tip, pendulum beam, and cart, respectively, $L$ is the length of the pendulum, $l=\frac{1}{2}L$ is the location of the beam center of mass, $k_1$ and $k_2$ are the spring constants of the pendulum tip and pendulum beam, respectively, $z$ is the cart motion, $\phi$ is the angular motion of the pendulum center of mass, $\theta$ is the angular motion of the pendulum tip, and $F$ is the force applied to the cart. Table \ref{tabapp} in the Appendix provides values for the model parameters. 

$$
    \mathcal{L} = \frac{1}{2} m_c \Dot{z}^2 + \frac{1}{2} m_b (\Dot{z}^2 _ 2 l \Dot{z} \Dot{\phi}cos\phi + l^2 \Dot{\phi})
$$
$$
    + \frac{1}{2} m_t [\Dot{z}^2 + 2 \Dot{z}(L \Dot{\theta} cos \theta + l \Dot{\phi} cos \phi + L^2 \Dot{\theta}^2
$$
$$
    + l^2 \Dot{\phi}^2 + 2 L l \Dot{\theta} \Dot{\phi} cos (\theta - \phi)] + m_t g L (1-cos\theta)
$$
\begin{equation}\label{lagrange}
    - \frac{1}{2} k_1 (L sin \theta)^2 + m_b g l (1-cos\phi) - \frac{1}{2} k_2 (l sin \phi)^2
\end{equation}

Structural damping of the pendulum as well as friction damping of the cart are modeled as Rayleigh damping, given in Equation \ref{damp} where $b_1$, $b_2$, and $b_3$ are the damping coefficients of the tip, center of mass, and cart, respectively.

$$
    R_d = \frac{1}{2} b_3 \Dot{z}^2 + \frac{1}{2} b_2 [(\Dot{z} + l \Dot{\phi} cos \phi)^2 + l^2 \Dot{\phi}^2 sin^2 \phi ] + 
$$

\begin{equation}\label{damp}
    \frac{1}{2} b_1 [(\Dot{z} + l \Dot{\phi} cos \phi + L \Dot{\theta} cos \theta)^2 + (l \Dot{\phi} sin \phi + L \Dot{\theta} sin \theta)^2]
\end{equation}

Equation \ref{euler} describes the Euler-Lagrange equation that when carried out with the Lagrangian given in Equations \ref{lagrange} and \ref{damp} produces the system equations of motion, where $w = \begin{bmatrix} \theta &  \phi & z \\ \end{bmatrix}$ and $F = \begin{bmatrix} 0 & 0 & F \\ \end{bmatrix}$. 

\begin{equation}\label{euler}
    \frac{\partial}{\partial t}\frac{\partial \mathcal{L}}{\partial \Dot{w}_i} - \frac{\partial \mathcal{L}}{\partial w_i} + \frac{\partial R_d}{\partial \Dot{w_i}} = F_i
\end{equation}

The system equations of motion are defined by Equations \ref{sys1}-\ref{sys3}.
$$
    m_t L^2 \Ddot{\theta} + m_t L \Ddot{z} cos \theta + m_t L l \Ddot{\phi} cos (\theta - \phi)
$$
$$
    + m_t L l \Dot{\phi}^2 sin (\theta - \phi) + \frac{1}{2} k_1 L^2 sin(2\theta) - m_t g L sin \theta
$$
\begin{equation}\label{sys1}
    b_1 L \Dot{z} cos \theta + b_1 L^2 \Dot{\theta} + b_1 L l \Dot{\phi} cos(\theta - \phi) = 0
\end{equation}

$$
    (m_t + m_b)l^2 \Ddot{\phi} + (m_t + m_b) l \Ddot{z} cos \phi + m_t L l \Ddot{\theta}cos(\theta-\phi)
$$
$$
    + (b_1 + b_2) l \Dot{z} cos \phi + (b_1+b_2) l^2 \Dot{\phi} + b_1 L l \Dot{\theta} cos (\theta-\phi)
$$
\begin{equation}\label{sys2}
    - m_t L l \Dot{\theta}^2 sin(\theta-\phi) + \frac{1}{2}k_2 l^2 sin(2\phi) - m_b g l sin \phi = 0
\end{equation}

$$
    (m_c+m_b+m_t) \Ddot{z} + (m_t + m_b) l (\Ddot{\phi} cos \phi - \Dot{\phi}^2)
$$
$$
    m_t L (\Ddot{\theta} cos\theta - \Dot{\theta}^2 sin\theta) + (b_1+b_2+b_3) \Dot{z}
$$
\begin{equation}\label{sys3}
    (b_1+b_2) l \Dot{\phi} cos \phi + b_1 L \Dot{\theta} cos \theta = F
\end{equation}

This system of nonlinear differential equations is transformed into a matrix format as shown in Equations \ref{statevec} and \ref{matsys}. The matrix equations are numerically integrated with the fourth order Runge-Kutta algorithm provided in the Python module GNCPy \cite{gncpy}. 

\begin{equation}\label{statevec}
    \Vec{x}=\begin{bmatrix}
    z & \Dot{z} & \phi & \Dot{\phi} & \theta & \Dot{\theta} \\
    \end{bmatrix}^T
\end{equation}

\begin{equation}\label{matsys}
    \Dot{\Vec{x}}=\underline{A}(\Vec{x})\Vec{x} + \underline{B}(\Vec{x}) F + \underline{L}(\Vec{x})
\end{equation}

Disturbances are injected into the system through the rate terms in $\Vec{x}$ and values are determined via three independent Ornstein-Uhlenbeck (OU) stochastic processes \cite{OU}. Equation \ref{OU} describes the OU process stochastic differential equation, where $y(t)$ is the disturbance magnitude, $\kappa$ is the rate of mean reversion, $\mu$ is the long-term mean of the process, $W(t)$ is the Brownian motion stochastic process, and $\sigma$ is the average magnitude of the Brownian motion.

\begin{equation}\label{OU}
    dy(t) = \kappa (\mu - y(t))dt + \sigma dW(t)
\end{equation}

\begin{figure}[htbp]
\centerline{\includegraphics[width=0.5\textwidth]{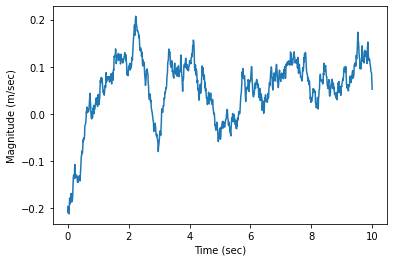}}
\caption{Sample time history of cart disturbance stochastic process.}
\label{discart}
\end{figure}

\begin{figure}[htbp]
\centerline{\includegraphics[width=0.5\textwidth]{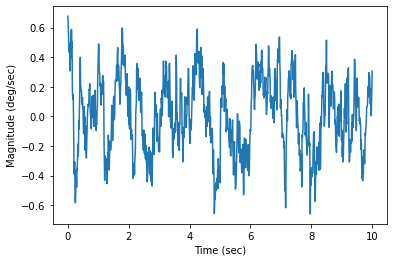}}
\caption{Sample time history of pendulum disturbance stochastic process.}
\label{dispend}
\end{figure}

At each time step, the three independent stochastic processes are sampled and added to the associated state value. Figures \ref{discart}-\ref{dispend} provide sample time histories of the OU processes for the cart velocity and pendulum angular velocities, respectively. OU process parameters for each disturbance are provided in the appendix. 

\subsection{Parametric Uncertainty Modeling}

Parametric uncertainty is applied to the model through the pendulum’s structural stiffness and damping coefficients and the cart’s friction damping coefficient, $k_{1,2}$ and $b_{1,2,3}$. At each instantiation of the Python class describing the model, these values are sampled from a Gaussian probability distribution with mean equal to the nominal value and covariance of 50\% of the nominal value.  This is applied to the five uncertain parameters prior to the start of each simulation run. 

To protect against negative stiffness or damping due to a sample far along the distribution tail, the absolute value of each parameter is used for mathematical model construction. This makes the probability density function of the uncertain parameters a folded normal distribution.

\subsection{Deep Reinforcement Learning Control}
Deep reinforcement learning algorithms operate by controlling an agent that, at each discrete time step $t$, performs actions $a$ that alter the state $s$ of an environment and receives rewards $r$. In this work, the agent is the control system, the environment is the FIPWC model, the actions are the control force applied to the cart, the states are the state vector given in Equation \ref{statevec}, and the reward is defined below. 

The behavior of the agent is defined by a policy $\pi$ which maps states to a probability distribution over the actions. The environment $E$ is stochastic and modeled at a Markov decision process with state space $\mathcal{S}$, action space $\mathcal{A}=\mathbb{R}^N$, an initial state distribution $p(s_1)$, transition dynamics $p(s_{t+1}|s_t, a_t$, and reward function $r(s_t, a_t)$. The initial state distribution is assumed to be zero $\Vec{x}=\Vec{0}$, the transition dynamics are given by the deterministic equations of motion above combined with the stochastic disturbances and parametric uncertainty, and the reward function is given by Equation \ref{reward} below. 

\begin{figure}[htbp]
\centerline{\includegraphics[width=0.5\textwidth]{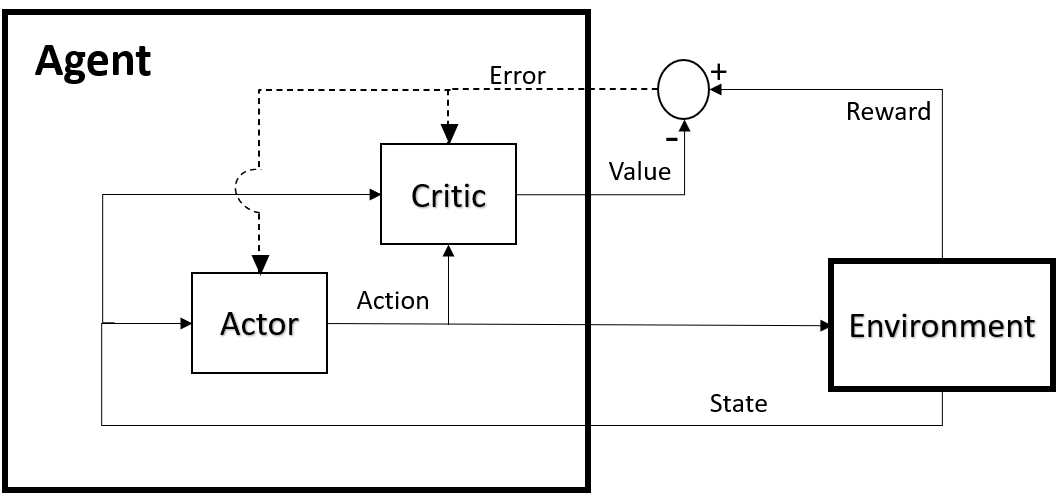}}
\caption{Block diagram for actor-critic agent.}
\label{drldiag}
\end{figure}

DDPG utilizes an actor-critic formulation with two neural network function approximators, one to choose what action to take and the other to decide how valuable that action was. Figure \ref{drldiag} provides a block diagram of the DDPG agent system.  

The critic function $Q(s, a)$ learns using the Bellman equation given in Equation \ref{bellman} where $\gamma$ is the discount factor $0<\gamma<1$ and $\alpha$ is the learning rate. A discount factor close to zero prioritizes near-term rewards, while a factor close to one places importance on long-term rewards. The learning rate is a critical parameter, and is typically kept low to ensure stable learning.

$$
    Q(s_{t+1}, a_{t+1}) = Q(s_t, a_t) + \alpha [r_t
$$
\begin{equation}\label{bellman}
     + \gamma \underset{a}{max} Q(s_{t+1}, a) - Q(s_t, a_t)]
\end{equation}

The actor function $\mu ( s | \theta^{\mu})$ specifies the current policy by mapping states to actions and is updated by applying the chain rule to the expected return from the initial distribution, $J=\mathbb{E}(r_1)$, with respect to the actor parameters as shown in Equation \ref{actor}. This is the gradient of the policy's performance \cite{silver}, also known as the policy gradient in DDPG.

\begin{equation}\label{actor}
    \nabla_{\theta, \mu} J = \mathbb{E}[\nabla_a Q(s, a|\theta^Q)|_{s=s_t, a=\mu(s_t)} \nabla_{\theta_\mu}\mu(s|\theta^{\mu})|_{s=s_t}]
\end{equation}

However, it is not advantageous to always choose the "best" action. At the beginning of a training session, the neural networks will be poorly optimized and will not know what the best actions are. Therefore $\mu(s_t | \theta_t^\mu )$ is augmented by additive noise sampled from a temporally correlated stochastic process, in this work from the OU process. As training progresses, the noise scaling term $\epsilon (t)$ is decreased to transition the agent from "exploration" of the state space to "exploitation" of what it considers to be good actions.

\begin{equation}
    \mu ' (s_t) = \mu(s_t | \theta_t^\mu ) + \epsilon (t) \mathcal{N}
\end{equation}

The reward function used in this work is given by Equation \ref{reward}, where $\Delta t$ is the simulation time step, $\Vec{x}_i$ is the $i^{th}$ term of the state vector, $\Vec{x}_{i,des}$ is the desired value for the $i^{th}$ state (each equal to zero for this application), $F$ is the control force, and $w_i$ is the reward tuning weight for the $i^{th}$ state. This function is arbitrary, and was chosen specifically for this application. Methods exist to learn the reward function, known as Inverse Reinforcement Learning (IRL), and it is possible to simultaneously learn the reward and policy functions through a coupled approach \cite{irlglmb}. 

\begin{equation}\label{reward}
    r = -\Delta t ( \sum_{i=1}^{N_s} w_i (\Vec{x}_i - \Vec{x}_{i,des})^2 + 0.1 F^2 )
\end{equation}

$$
    w =\begin{bmatrix}\label{weightvec}
    0.1 & 0.5 & 1 & 1 & 1.2 & 1 \\
    \end{bmatrix}
$$

This reward function incentivizes keeping the system state close to the desired value, in this case with the pendulum vertical, the cart near its initial horizontal position, and the time derivatives small. The control effort reward is scaled by 0.1 to ensure that the agent learns that while it is important to keep the control effort low, it is much more critical to keep the states close to their desired values. This sum is then multiplied by the negative of the simulation time increment to both scale the reward and transform from a cost function (to be minimized) to a reward function (to be maximized). 

This work utilizes the DDPGAgent Python class provided by the open-source module Keras-RL \cite{kerasrl}, neural network functionality from TensorFlow \cite{tensorflow}, and a custom OpenAI Gym environment \cite{gym} to cohere the simulation model to Keras-RL. The actor neural network uses twenty hidden layers with 256 neurons per layer, while the critic neural network uses the same number of hidden layers with 512 neurons per layer. The agent was trained for 100,000 steps with 512 batches per step. Learning parameters are given in Table \ref{tabapp} in the appendix.

\section{Simulation Results}

Due to the stochastic nature of this work, a great number of simulations utilizing the Monte Carlo approach were conducted to evaluate control system performance. 10,000 simulations for both the classical proportional-derivative (PD) and deep reinforcement learning control systems were performed with rewards calculated per Equation \ref{reward}, and the average reward and the standard deviation of reward were extracted. Additionally, 10,000 simulations per control system were performed without the cart velocity disturbance $d_{\Dot{z}}$ as the system, particularly with PD control, was notably sensitive to this disturbance. These results are summarized in Table \ref{tab1}.

\begin{table}[h]
\renewcommand{\arraystretch}{1.3}
\caption{Reward Statistics from 10,000 Simulations}
\label{table_example}
\centering
\begin{tabular}{|c|c|c|}
\hline
\textbf{Control System} & \textbf{Average} & \textbf{Standard Deviation} \\
\hline
PD no $d_{\Dot{z}}$ & -111.72 & 183.52 \\
\hline
DRL no $d_{\Dot{z}}$ & -63.83 & 61.12 \\
\hline
PD & -79,005.72 & 132,957.99 \\
\hline
DRL & -212.35 & 105.79 \\
\hline
\end{tabular}
\label{tab1}
\end{table}

\begin{figure}[htbp]
\centerline{\includegraphics[width=0.5\textwidth]{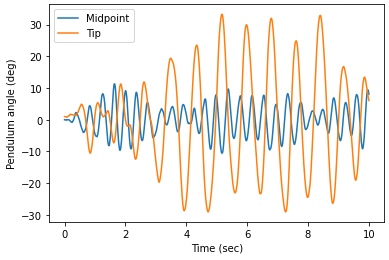}}
\caption{Example time history of pendulum angles for PD control without $d_{\Dot{z}}$.}
\label{pdang}
\end{figure}

\begin{figure}[htbp]
\centerline{\includegraphics[width=0.5\textwidth]{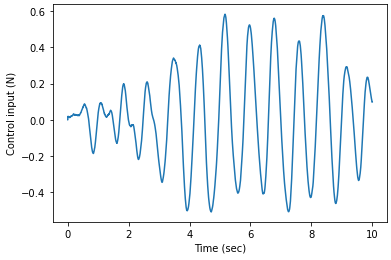}}
\caption{Example time history of control force for PD control without $d_{\Dot{z}}$.}
\label{pdcon}
\end{figure}

\begin{figure}[htbp]
\centerline{\includegraphics[width=0.5\textwidth]{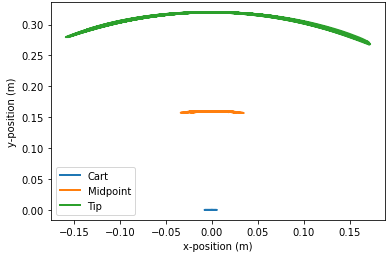}}
\caption{Example FIPWC motion for PD control without $d_{\Dot{z}}$.}
\label{pdpen}
\end{figure}

\begin{figure}[htbp]
\centerline{\includegraphics[width=0.5\textwidth]{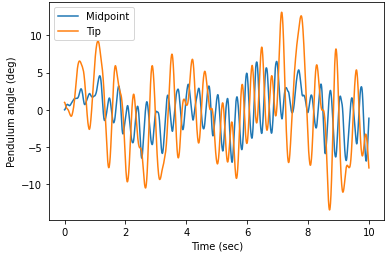}}
\caption{Example time history of pendulum angles for DRL control with $d_{\Dot{z}}$.}
\label{drlang}
\end{figure}

\begin{figure}[htbp]
\centerline{\includegraphics[width=0.5\textwidth]{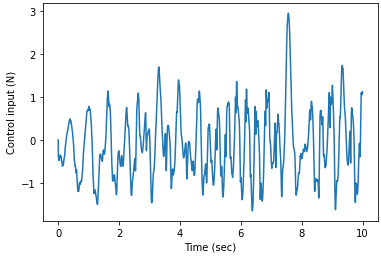}}
\caption{Example time history of control force for DRL control with $d_{\Dot{z}}$.}
\label{drlcon}
\end{figure}

\begin{figure}[htbp]
\centerline{\includegraphics[width=0.5\textwidth]{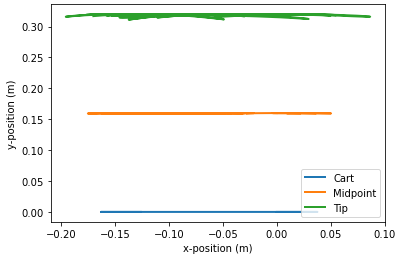}}
\caption{Example FIPWC motion for DRL control with $d_{\Dot{z}}$.}
\label{drlpen}
\end{figure}

\begin{figure}[htbp]
\centerline{\includegraphics[width=0.5\textwidth]{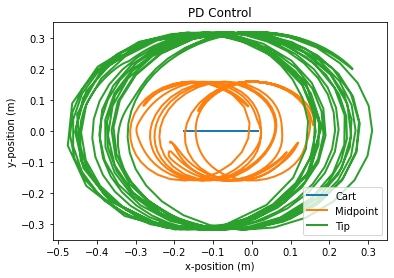}}
\caption{Example FIPWC motion for PD control with $d_{\Dot{z}}$.}
\label{pdbad}
\end{figure}

Figures \ref{pdang}-\ref{pdbad} present data from representative simulations for both control systems. Figures \ref{pdang} and \ref{drlang} show the time history of the pendulum midpoint and tip angles, Figures \ref{pdcon} and \ref{drlcon} plot the time history of the control force applied to the cart, and Figures \ref{pdpen} and \ref{drlpen} display the two-dimensional motion of the cart, pendulum midpoint, and pendulum tip. Figure \ref{pdbad} shows a typical simulation of the PD system when $d_{\Dot{z}}$ is present. Note that $d_{\Dot{z}}$ is present for all DRL plots.

\section{Discussion}

It is clear from Table \ref{tab1} that the DRL control system is able to keep the states closer to their desired values, in this case representing the pendulum vertical, the cart close to its initial position, and the rates low, for longer compared to the PD control system, particularly when $d_{\Dot{z}}$ is present. The PD controller is wholly incapable of rejecting $d_{\Dot{z}}$, while the DRL controller is able to stabilize the system for all combinations of disturbance levels as indicated by the low standard deviations in Table \ref{tab1}. The DRL controller achieves nearly double the reward average and half the standard deviation of the PD controller even when $d_{\Dot{z}}$ is neglected since the system is regularly driven outside of the linear regime, commonly considered to be $| \phi, \, \theta| < 15^{\circ}$.

Figures \ref{pdang}-\ref{pdbad} provide a visual insight into the superior performance of the DRL system. The simulations that produced these data represent a typical result close to the average performance for each system. Despite the DRL controller having to compensate for $d_{\Dot{z}}$, the pendulum angles are kept below 11 degrees for the entire simulation and below 10 degrees for two thirds of the run. The sharp increase in the angles from 7-9 seconds can be attributed to a particularly high-magnitude combination of disturbances, with the controller beginning to compensate for them as the simulation ends. 

Figure \ref{pdbad} demonstrates the PD control system's difficulties in keeping the pendulum vertical with $d_{\Dot{z}}$ present. The standard deviation for this set of runs is particularly high because in many simulations the angular rates reached the DDPG agent's observation space limit of $\pm 1 \times 10^9$, which returns $-1 \times 10^7$ for the reward. This severely punishes the agent for exploring this region of the state space, as the OpenAI Gym environment returns $nan$ when the state is outside of that space which spoils any training or testing runs.

\section{Conclusions}
This work presented a deep reinforcement learning control approach to stochastic disturbance rejection for a highly nonlinear dynamic system with parametric uncertainty. Results from 10,000 Monte Carlo simulations demonstrated superior performance for the DRL control system compared to a classical proportional-derivative control law. 

While the DRL controller produced good results, there are still some improvements that could be made. The controller is operating at the simulation rate of 100 Hz, and has no rate limitation as is clear from Figure \ref{drlcon}. This is not very realistic since real actuators would not operate at such a high frequency, have rate limitations, and be subject to their own dynamic effects such as rise time, settling time, and overshoot. The DDPGAgent class has a parameter to decrease the policy update rate from the simulation rate. This could be used to limit the controller rate to 10 Hz, for example. An actuator dynamics model could also be incorporated to more accurately represent real-world systems. 

Future work topics of interest are implementing realistic disturbances, nonlinear flight vehicle models, and coupling the DDPG agent with a recurrent neural network (RNN) for uncertain parameter estimation and disturbance observation.

\section*{Appendix}

Table \ref{tabapp} provides the simulation parameter values used in this work.

\begin{table}[h!]
\renewcommand{\arraystretch}{1.3}
\caption{Simulation Parameter Values}
\label{table_example}
\centering
\begin{tabular}{|c|c||c|c|}
\hline
\textbf{Parameter} & \textbf{Value (unit)} & \textbf{Parameter} & \textbf{Value (unit)} \\
\hline
$m_t$ & 0.019 kg & $m_c$ & 0.18 kg \\
\hline
$m_b$ & 0.0215 kg & $L$ & 0.32 m \\
\hline
$l$ & 0.16 m & $b_1$ & 0.001 N-s/m \\
\hline
$b_2$ & 0.001 N-s/m & $b_3$ & 12 N-s/m \\
\hline
$k_1$ & 2 N/m & $k_2$ & 8 N/m \\
\hline
$\Delta t$ & 0.01 sec & $g$ & 9.81 $m/s^2$ \\
\hline
$\kappa_{\Dot{z}}$ & 0.01 & $\kappa_{\Dot{\phi}, \Dot{\theta}}$ & 10 \\
\hline
$\mu_{\Dot{z}}$ & 0 & $\mu_{\Dot{\phi}, \Dot{\theta}}$ & 0 \\
\hline
$\sigma_{\Dot{z}}$ & 0.1 m/s & $\sigma_{\Dot{\phi}, \Dot{\theta}}$ & 1 deg/sec \\
\hline
$\alpha$ & 0.001 & $\gamma$ & 0.99 \\

\hline
\end{tabular}
\label{tabapp}
\end{table}

\end{document}